# Impact of Quantum Well Thickness on Efficiency Loss in InGaN/GaN LEDs: Challenges for Thin-Well Designs

Xuefeng Li[1*], Nick Pant[2,3], Sheikh Ifatur Rahman[4], Rob Armitage[5], Siddharth Rajan[4,6], Emmanouil Kioupakis[2], and Daniel Feezell[1]

[1]Center for High Technology Materials (CHTM), University of New Mexico, Albuquerque, NM 87106, USA
[2]Department of Materials Science & Engineering, University of Michigan, Ann Arbor, MI 48109, USA
[3]Applied Physics Program, University of Michigan, Ann Arbor, MI 48109, USA
[4]Department of Electrical and Computer Engineering, The Ohio State University, Columbus, OH 43210, USA
[5]Lumileds LLC, San Jose, CA 95131, USA
[6]Department of Materials Science and Engineering, The Ohio State University, Columbus, OH 43210, USA

We investigate the impact of quantum well (QW) thickness on efficiency loss in *c*-plane InGaN/GaN LEDs using a small-signal electroluminescence (SSEL) technique. Multiple mechanisms related to efficiency loss are independently examined, including injection efficiency, carrier density vs. current density relationship, phase space filling (PSF), quantum confined stark effect (QCSE), and Coulomb enhancement. An optimal QW thickness of around 2.7 nm in these InGaN/GaN LEDs was determined for quantum wells having constant In composition. Despite better control of deep-level defects and lower carrier density at a given current density, LEDs with thin QWs still suffer from an imbalance of enhancement effects on the radiative and intrinsic Auger-Meitner recombination coefficients. The imbalance of enhancement effects results in a decline in internal quantum efficiency (IQE) and radiative efficiency with decreasing QW thickness at low current density in LEDs with QW thicknesses below 2.7 nm. We also investigate how LED modulation bandwidth varies with quantum well thickness, identifying the key trends and their implications for device performance.

___________________________

[*]**Electronic mail:** xuefengli@ucsb.edu

The significant efficiency loss in green InGaN/GaN LEDs is a challenge for solid-state lighting and is often referred to as the "green gap" [1]. The emission wavelength of an LED is primarily determined by the quantum well (QW) thickness and the indium composition. Our recent study demonstrated that commercial-grade LEDs with 3-nm-thick QWs, but with different indium compositions, do not exhibit a significant decrease in growth quality with higher indium composition. However, a strong increase in carrier density was observed at a given current density for longer wavelength LEDs [2], leading to efficiency loss at higher indium compositions. Furthermore, we also reported that the effect of deep-level defect-related nonradiative recombination is very limited in 3 nm InGaN/GaN LEDs with commercial-grade epitaxy, while the intrinsic Auger-Meitner process remains the dominant nonradiative mechanism associated with the green gap [3]. Along with indium composition and defect density, QW thickness is also a crucial parameter that impacts quantum efficiency in InGaN/GaN LEDs. In commercial LEDs, QW thicknesses greater than 3.0 nm are usually avoided due to the high density of deep-level defects introduced by the large lattice mismatch between InGaN and GaN layers [4,5]. In addition, thick QW LEDs exhibit strong quantum-confined stark effect (QCSE) and small wavefunction overlap [6-9], which have been shown to significantly increase the carrier density and therefore exacerbate intrinsic Auger-Meitner recombination [10-13]. The use of thin QWs in InGaN/GaN LEDs is known to help to mitigate QCSE and may provide benefits by lowering the carrier density, but such designs require a higher indium composition for a given wavelength compared to designs with thicker QWs and may face other limitations. Thus, a systematic study of the effect of QW thickness on carrier behavior and efficiency loss in commercial-grade LEDs is essential for identifying the limiting factors and providing practical solutions to increase efficiency in InGaN/GaN LEDs.

In this work, we study a series of *c*-plane InGaN/GaN LEDs with various QW thicknesses, grown with state-of-the-art growth conditions using metal-organic chemical vapor deposition (MOCVD). The LED series includes six wafers with identical indium compositions (19% based on energy-dispersive x-ray spectroscopy (EDX) measurements), but different QW thickness that range from 2.1 nm to 3.6 nm. The barriers consist of 18-nm-thick GaN layers. Although the LEDs have 3X, 3-nm-thick QWs, the carrier recombination is limited to a single QW next to the p-GaN [2, 3]. Detailed information on the emission wavelengths and carrier densities at 40 A/cm$^2$ for the LED series is given in Table 1.

Small-signal electroluminescence (SSEL) [14, 15] was used to study the carrier dynamics in the LEDs in Table 1. SSEL is a technique based on electrical injection that analyzes the impedance and modulation response of LEDs. Details regarding the SSEL setup and process can be found in Ref. 2 and 15. Multiple carrier dynamics parameters related to carrier recombination and transport can be acquired using SSEL, including carrier density, lifetime, and recombination rate. In the SSEL circuit model utilized in this work, differential injection efficiency ($\eta_{\Delta inj}$) is defined as the ratio of carriers that recombine in the active region, rather than carriers injected into the active region, to the total carrier injection in the LED. Injection efficiency ($\eta_{inj}$) can be acquired via integrating $\eta_{\Delta inj}$ [16].

| QW thickness [nm] | Wavelength @ 40 A/cm$^2$ [nm] | $n$ @ 40 A/cm$^2$ [cm$^{-3}$] |
| --- | --- | --- |
| 2.1 | 465 | $4.57 \times 10^{18}$ |
| 2.4 | 487 | $7.83 \times 10^{18}$ |
| 2.7 | 504 | $1.10 \times 10^{19}$ |
| 3.0 | 523 | $1.68 \times 10^{19}$ |
| 3.3 | 539 | $2.46 \times 10^{19}$ |
| 3.6 | 570 | $4.79 \times 10^{19}$ |

Table 1. QW thickness, emission wavelength, and $n$ @ 40 A/cm$^2$ in the LED series.

As shown in Figure 1(a), the injection efficiency is higher in LEDs with thinner QWs, and only small differences are observed at high current densities. The low injection efficiency in LEDs with thicker QWs results from relatively slower carrier recombination rates in the active region and increased carrier recombination in the cladding layer [16]. The internal quantum efficiency (IQE) in Figure 1(b) was obtained from Lumileds by measuring the external quantum efficiency (EQE) of LEDs with known light extraction efficiency (LEE). The IQEs maintain similar values for QW thickness of 3.0 nm and lower but decrease notably for QW thickness of 3.3 nm and 3.6 nm, especially at low current density. The radiative efficiency ($\eta_r$) was derived from the IQE ($\eta_{IQE}$) and injection efficiency ($\eta_{inj}$) using $\eta_{IQE} = \eta_{inj} * \eta_r$, and is shown in Figure 1(c). Similar trends were observed in radiative efficiency as in IQE. The significant decline in IQE and radiative efficiency in LEDs with QW thickness greater than 3.3 nm is likely due to higher densities of deep-level defects in thick QWs.

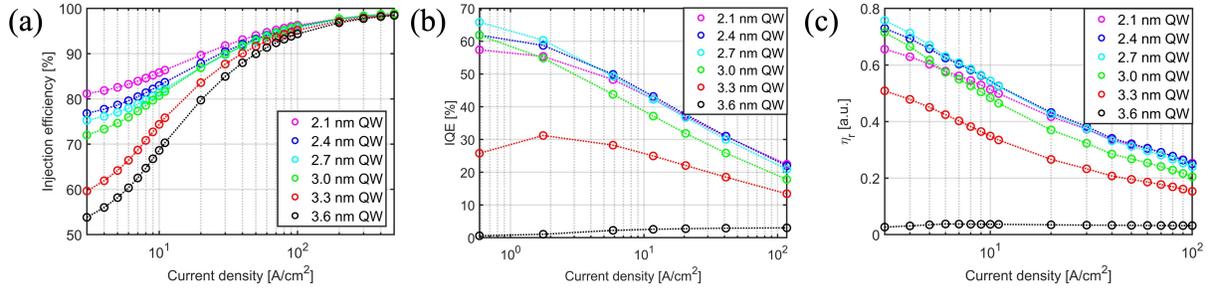

FIG. 1. (a) Injection efficiency, (b) IQE, and (c) Radiative efficiency vs. $J$ in LEDs with different QW thickness.

The relationship between carrier density ($n$) and current density ($J$) is an important factor in the efficiency of InGaN/GaN LEDs since it not only determines how carriers are converted into current, but also affects the relative contributions of the various recombination mechanisms at a given $J$. Maintaining a lower $n$ is advantageous for avoiding strong intrinsic Auger-Meitner recombination and for preventing significant efficiency loss at high $J$. The results for the $n$ vs. $J$ relationship [16] are shown in Figure 2(a). LEDs with thinner QWs have higher recombination coefficients due to weaker QCSE and larger wavefunction overlap. Thinner QWs maintain lower carrier density, indicating that the effect of a thinner QW is not the dominant factor that increases $n$. Vice versa also holds, where thicker QWs do not necessarily lead to smaller $n$. Instead, the increase in recombination coefficients and decrease in carrier lifetime are the dominant factors affecting carrier density. The $n$ values at 40 A/cm$^2$ in the QW thickness LED series are shown in Table 1. Notably, the corresponding $n$ value at 40 A/cm$^2$ increases by 10.5X when the QW thickness increases from 2.1 nm to 3.6 nm. A similar analysis of the $n$ vs. $J$ relationship for other LED series can be found in Ref. 2 and 3.

In Figure 2(b), we plot $\eta_r$ vs. $n$. LEDs with thin QWs have lower radiative efficiency at a given $n$. The decline in radiative efficiency at low $J$ in LEDs with QW thicknesses below 2.7 nm, as shown in Figure 1(c), correlates with a more significant difference at low $n$ in the $\eta_r$ vs. $n$ plot, as shown in Figure 2(b). In Figures 2(c) & (d) and 2(e) & (f), we show the radiative and nonradiative recombination lifetimes ($\tau_r$ and $\tau_{nr}$) and rates ($R_r$ and $R_{nr}$), respectively. LEDs with thin QWs have shorter recombination lifetimes and faster recombination rates, which are related to the larger wavefunction overlap and other enhancement mechanisms, which will be discussed in the next section.

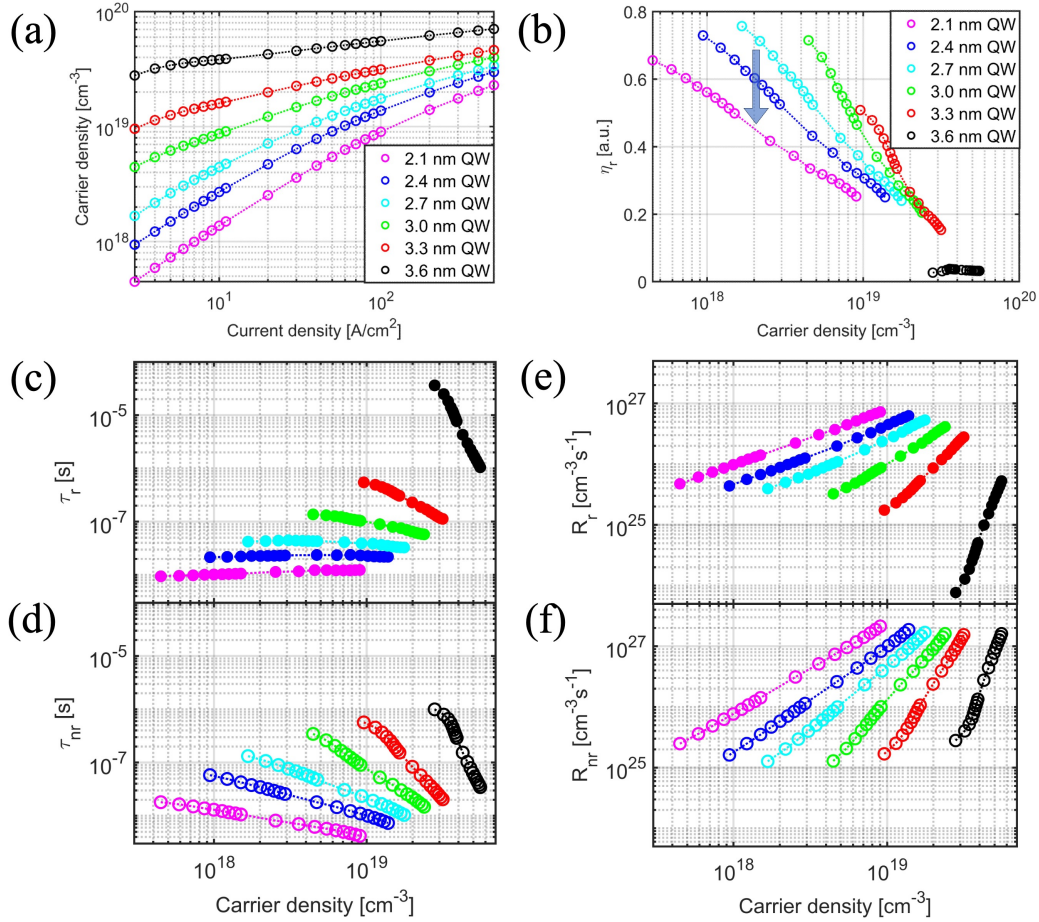

FIG. 2. (a) Carrier density ($n$) vs. current density ($J$), (b) Radiative efficiency vs. $n$, (c) & (d) Radiative and nonradiative lifetimes vs. $n$, and (e) & (f) Radiative and nonradiative recombination rates vs. $n$ in LEDs with different QW thickness.

The recombination coefficients, $A(n), B(n)$, and $C(n)$ can be acquired by plotting $R_{nr}/n$, $R_r/n^2$, and $(R_{nr} - An)/n^3$, respectively, as shown in Figure 3 [17]. Here, $A$ is the value of $A(n)$ at low $n$. Although higher deep-level defect densities are expected in thick QW LEDs, $R_{nr}/n$ remains lower compared to thin-QW LEDs. This discrepancy can be attributed to the strong QCSE and small wavefunction overlap in thick QWs [6-9]. We bracket the range of the $A$ parameter since $R_{nr}/n$ doesn't converge to a flat trend at low $n$ in the carrier density range measured here. The $B(n)$ coefficient is derived from $R_r/n^2$ and shows strong enhancement at low $n$ in LEDs with QW thickness below 3.0 nm. This enhancement is typically attributed to carrier localization and Coulomb enhancement in thin QWs, characterized by strong internal electrical field and localized carriers [2, 3, 18]. In the 3.3 nm QW LED, limited enhancement of the $B(n)$ coefficient was observed at low $n$, accompanied by some screening effect at high $n$. The 3.6-nm-QW LED shows no enhancement of the $B(n)$ coefficient at low $n$ but shows a strong screening effect at high $n$. The $C(n)$ coefficient is derived from $(R_{nr} - An)/n^3$ and shows a strong enhancement at low $n$ in LEDs with QW thickness less than 2.7 nm. In Figure 3, the spread in the $C(n)$ coefficient values result from the previously discussed bracketed range of $A$ values. A significant enhancement of the $C(n)$ coefficient was observed at low $n$ in QWs with thickness below 2.7 nm, consistent with a similar behavior reported in 2.5 nm QWs [19]. To the best of our knowledge, no other enhancement effects on recombination coefficients at low $n$ have been reported, except for the effect of carrier localization and Coulomb enhancement on the $B(n)$ coefficient in thin-QW LEDs [18]. Further work is needed to determine why $C(n)$ is enhanced at low $n$. In our view, a likely explanation is the many-body Coulomb enhancement due to electron-hole correlations [20, 21], which correspond to composite electron-electron-hole or hole-hole-electron quasiparticles. This interaction is suppressed by the polarization field for thicker wells. However, a theoretical analysis of tronic Coulomb enhancement effects on the $C(n)$ coefficient is challenging due to the complexity of the simulation and has not been reported. If carrier localization and Coulomb enhancement were to primarily influence the $B(n)$ coefficient and not the $C(n)$ coefficient for some QW thicknesses, we would anticipate a rise in radiative efficiency at low $J$. Notably, while an enhancement effect on $B(n)$ was observed in the 3.0 nm QW LED, no such effect was observed on $C(n)$, indicating that a QW thinner than 3.0 nm may be required to observe the enhancement effect on $C(n)$.

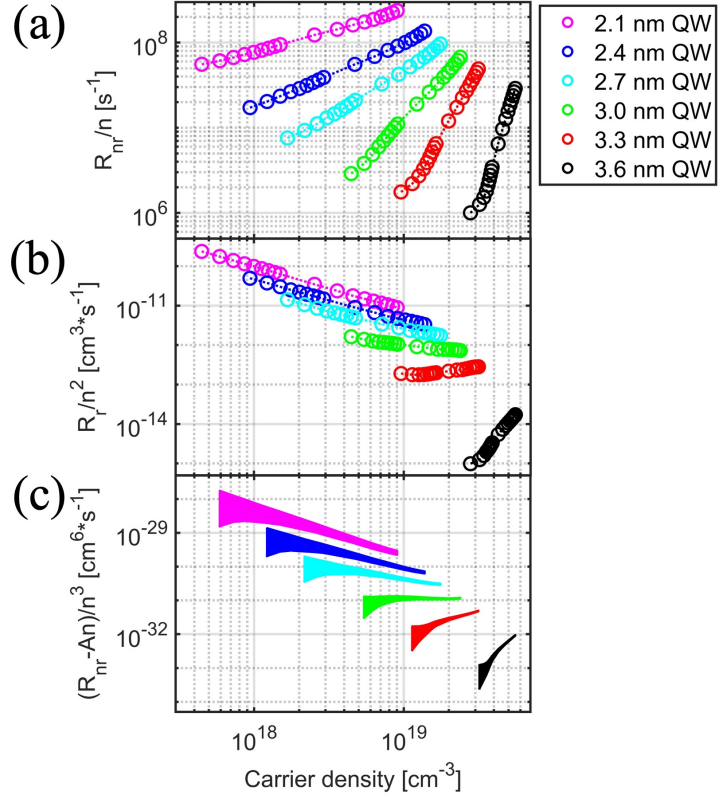

FIG. 3. (a)-(c) $R_{nr}/n$, $R_r/n^2$, and $(R_{nr} - An)/n^3$ in LEDs with different QW thickness.

We previously showed that the influence of deep-level defect-related recombination mechanisms is small within the studied $J$ range in similar commercial-grade LEDs with 3 nm QWs [3]. Hence, the radiative efficiency at high $J$ can be approximated as:

$$\eta_r = \frac{B(n)n^2}{B(n)n^2 + C(n)n^3} = \frac{1}{1 + \frac{C(n)}{B(n)}n} \qquad (1)$$

There are only two factors that contribute to the radiative efficiency in the high $J$ range: $C(n)/B(n)$ and $n$. As shown in Figure 1(c), the radiative efficiencies for 2.1 nm, 2.4 nm, and 2.7 nm QW LEDs are nearly identical at high $J$. Since $n$ is lower in LEDs with thin QWs, as shown in Figure 2(a), the ratio $C(n)/B(n)$ is higher in LEDs with thin QWs for a given radiative efficiency at high $J$.

With the recombination coefficients obtained in Figure 3, we now calculate $C(n)/B(n)$, as shown in Figure 4. Due to the strong screening effect observed in 3.6 nm LED, the corresponding values of $C(n)$ shown in Figure 3(c) are less accurate, and we do not calculate $C(n)/B(n)$ for that LED. Here, we observe that LEDs with thinner QWs have higher values of $C(n)/B(n)$, consistent with the analysis related to equation 1. Additionally, we observe a different trend in $C(n)/B(n)$ vs. $n$ for LEDs with different QW thickness. For example, in the 3.0 nm QW LED, $C(n)/B(n)$ increases slightly at higher $n$. Conversely, $C(n)/B(n)$ shows a decreasing trend at a higher $n$ in the 2.1 nm QW LED.

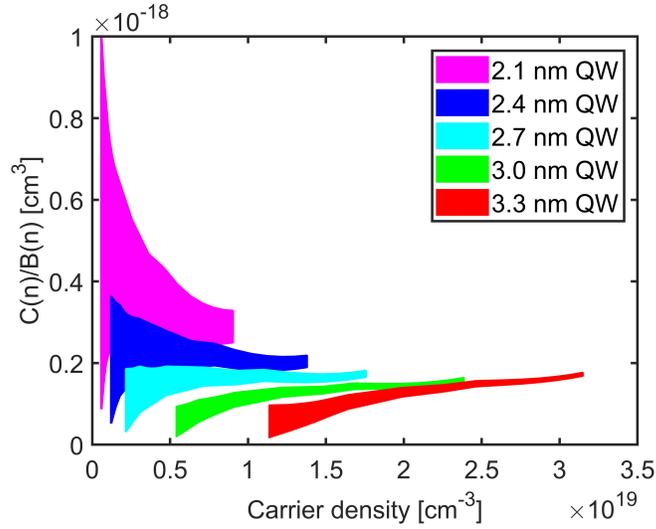

FIG. 4. $C(n)/B(n)$ vs. $n$ in LEDs with QW thickness of 2.1 nm, 2.4 nm, 2.7 nm, 3.0 nm, and 3.3 nm.

QW thickness also affects the -3dB bandwidth in the LEDs series, as shown in Figure 5(a). Generally, LEDs with thin QWs have wider -3dB bandwidths. Notably, there is an increase in the -3dB bandwidth with decreasing $J$ in thin-QW LEDs, such as in the 2.1 nm and 2.4 nm QW LEDs. The same behavior has been observed in thin-QW high-speed LEDs [22]. We can investigate the impact of carrier recombination and escape on the bandwidth to gain insight into the bandwidth trends. As the effect of parasitic resistance is small [23], the modulation response ($S_{21}$) can be expressed as:

$$S_{21} = -10 \log_{10} \left( 1 + \omega^2 \tau_{\Delta rec}^2 \frac{1}{\left(1 + \frac{\tau_{\Delta rec}}{\tau_{\Delta esc}}\right)^2} \right) + constant \tag{2}$$

Here, $\omega$ is the driving frequency. Then, $S_{21}(\omega_{3dB}) = S_{21}(\omega = 0) - 3dB$ and the modulation bandwidth ($\omega_{3dB}$) of the device is:

$$\omega_{3dB} = \frac{1}{\tau_{\Delta rec}} + \frac{1}{\tau_{\Delta esc}} = \frac{1 + \frac{\tau_{\Delta rec}}{\tau_{\Delta esc}}}{\tau_{\Delta rec}} \qquad (3)$$

Therefore, the modulation bandwidth is primarily determined by the differential recombination lifetime ($\Delta\tau_{rec}$), as shown in Figure 5(b), and the ratio between the differential recombination and escape lifetimes ($\frac{\tau_{\Delta rec}}{\tau_{\Delta esc}}$), as shown in Figure 5(c). $\tau_{\Delta rec}$ is relatively constant in thin QWs at low $J$ due to the strong enhancement effect and weak QCSE, leading to an increase in the modulation bandwidth of thin-QW LEDs at low $J$ when $\frac{\tau_{\Delta rec}}{\tau_{\Delta esc}}$ increases. For example, when $\frac{\tau_{\Delta rec}}{\tau_{\Delta esc}}$ is comparable to or larger than 1 (below ~10 A/cm²), the modulation bandwidth will be increased in the thin QWs. Thus, the interplay between carrier recombination and escape in thin QWs produces the trend in Figure 5(a). Strong carrier escape at low $J$ has been previously reported and attributed to weak Coulomb-enhanced capture into the QWs [14]. Similar behavior has also been observed in semipolar LEDs [16]. We do not observe an increase in modulation bandwidth at low $J$ for thick-QW LEDs because the modulation bandwidth behavior is dominated by a strong increase in $\tau_{\Delta rec}$ at low $J$ as described by equation 3 and shown in Figure 5(b).

We can gain further insight into radiative recombination by plotting the quantity $B(n)n$. As shown in Figure 5(d), this quantity shows a decreasing trend with increasing QW thickness and remains relatively stable with changes in $n$ in thin QWs. The fact that $B(n)n$ is stable at low $n$ for thin quantum wells indicates monomolecular recombination, which can be a signature of excitons, related to the Coulomb enhancement effect. However, some studies show an increase followed by a decrease in quantum efficiency with increasing $J$ in similar thin QWs [19, 24], which is difficult to explain with an excitonic model. Additionally, the properties of excitons in disordered InGaN/GaN QWs were numerically investigated, and carriers with excitonic behavior account for less than 25% of the total carriers in 2-nm-thick QWs [25]. In thick QW LEDs, $B(n)n$ declines at low $n$ because the enhancement of the $B(n)$ coefficient is limited, owing to suppression of the Coulomb enhancement by polarization fields.

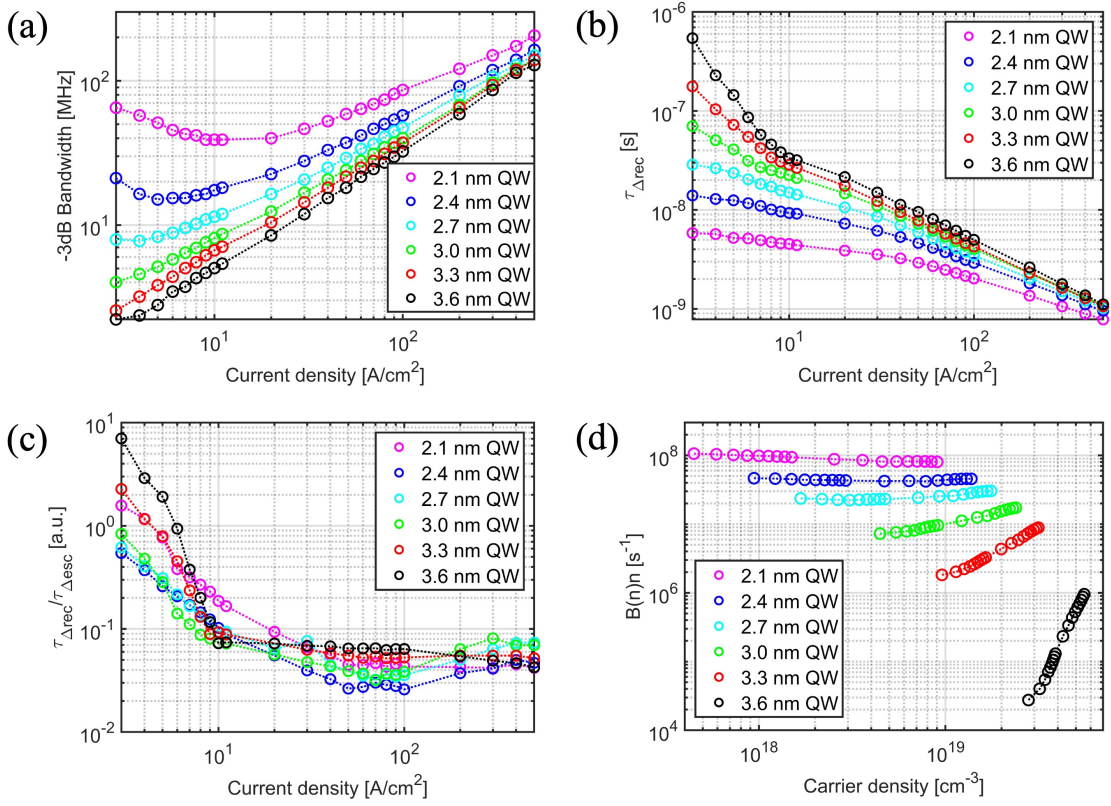

FIG. 5. (a) -3dB bandwidth, (b) $\tau_{\Delta rec}$, (c) $\tau_{\Delta rec}/\tau_{\Delta esc}$, and (d) $B(n)n$ in LEDs with different QW thicknesses.

As discussed earlier, we also observe a lower IQE and radiative efficiency with decreasing QW thickness below 2.7 nm in the low $J$ region, as shown in Figure 1(b) and 1(c). To gain insight into this, we focus the following analysis only on the low $n$ range. There are multiple mechanisms that affect the recombination coefficients, such as phase space filling (PSF) [17], QCSE [6-9], carrier localization [1, 26-31], and Coulomb enhancement [2, 18]. Notably, the PSF effect is weak at low $n$, especially in LEDs with thin QWs [17, 32], thus it was excluded from the low $n$ analysis. Additionally, the influence of trap-assisted Auger-Meitner processes on the radiative efficiency is negligible at low $n$ in LEDs with state-of-the-art growth conditions [3]. Therefore, at low $n$, the radiative efficiency can be expressed as:

$$\eta_r = \frac{B_0(n)n^2 I^b(n)}{A_0(n)nI^a(n) + B_0(n)n^2 I^b(n) + C_0(n)n^3 I^c(n)} \qquad (4)$$

Here, $I(n)$ represents wavefunction overlap in LEDs, and $a, b,$ and $c$ account for the dependence of SRH, radiative, and intrinsic Auger-Meitner processes on wavefunction overlap. $B_0(n)$ and $C_0(n)$ only consider the effects of carrier localization and Coulomb enhancement, excluding QCSE (which is accounted for in $I(n)$), on radiative and intrinsic

Auger-Meitner processes. $A_0(n)$ also excludes the effect of QCSE. After simplifying equation 4, the radiative efficiency can be written as:

$$\eta_r = \frac{1}{1 + \frac{A_0(n)I^{a-b}(n)}{B_0(n)n} + \frac{C_0(n)nI^{c-b}(n)}{B_0(n)}} = \frac{1}{1 + \frac{A_0(n)I^{a-b}(n)}{B_0(n)n} + n\frac{C_0(n)}{B_0(n)}} \quad (5)$$

The effect of wavefunction overlap is similar on both radiative and intrinsic Auger-Meitner processes, i.e., $b \approx c$ [2, 7, 31]. Therefore, $I^{c-b}(n) \approx 1$.

We now investigate why decreasing the QW thickness decreases the radiative efficiency at a given $n$, as shown by the arrow in Figure 2(b). In a previous work, we showed that defect-mediated recombination is small in commercial-grade LEDs with 3 nm quantum wells [2,3]. If we assume that this conclusion also holds for thinner quantum wells, then the decline in efficiency with decreasing thickness could be solely attributed to an increase in $C(n)/B(n)$, as shown in Figure 4. One explanation for this is well-width fluctuations, which are always present even in commercial-grade quantum wells. Such fluctuations have a stronger influence on the electronic structure of thinner wells, which could exacerbate $C(n)$ by relaxing crystal-momentum conservation. Although there is no experimental evidence from the present work, defect-mediated recombination, corresponding to the $\frac{A_0(n)I^{a-b}(n)}{B_0(n)n}$ term in equation 5, could also significantly impact the quantum efficiency in thin QWs. While David et al. have shown that the scaling exponent $a - b$ is approximately -0.1 for certain high quality quantum wells in the 2.5 nm to 5 nm thickness range [32], the scaling exponent can depend on the precise distribution of defects. If defects are clustered near the interface where electrons are confined by the polarization field, one might even observe superlinear scaling, where the defect-mediated term increases more dramatically than radiative recombination as the quantum well thickness decreases. (see supplementary discussion) The term $A_0(n)$, which depends on the multi-phonon-mediated capture coefficient can also have a non-trivial dependence on the quantum-well thickness as polarization fields can non-trivially influence the energetics of trap levels, i.e., shift a shallow defect into the mid-gap or vice versa. (see supplementary Figure S1) To our knowledge, this effect has not been previously considered, and further research is needed to determine whether the decline in radiative efficiency for thinner wells is an intrinsic limitation, or an extrinsic issue that can be mitigated through novel design paradigms.

In summary, we studied carrier dynamics in a series of InGaN/GaN LEDs with varying QW thickness ranging from 2.1 nm to 3.6 nm using SSEL. The optimal QW thickness for LED efficiency in this study is around 2.7 nm, which provides a good balance among injection efficiency, deep-level defect density, carrier-current density relationship, and carrier localization and Coulomb enhancement. As QW thickness increases, injection efficiency decreases, the carrier density increases radically, and likely additional deep-level defects are introduced during growth. Conversely, we observed a decline in IQE and radiative efficiency in the thinnest of the studied QWs, along with an increase in -3dB bandwidth with decreasing $J$, both in a similar low $J$ region in LEDs with thin QWs (less than 2.7 nm). In the present work, the data show that the imbalance of enhancement effects on the $B(n)$ and $C(n)$ coefficients and the influence of carrier escape are the primary reasons for the efficiency reduction and bandwidth trends observed in these thin-QW LEDs at low current density. Future work should also consider the potential role of defect-mediated recombination in thin QWs.

See the supplementary material for scaling exponent of SRH and radiative recombination.

This work was supported by the Department of Energy under Award No. DE-EE0009163.

# Impact of Quantum Well Thickness on Efficiency Loss in InGaN/GaN LEDs: Challenges for Thin-Well Designs (Supplementary Material)

Xuefeng Li[1*], Nick Pant[2,3], Sheikh Ifatur Rahman[4], Rob Armitage[5], Siddharth Rajan[4,6], Emmanouil Kioupakis[2], and Daniel Feezell[1]

[1]*Center for High Technology Materials (CHTM), University of New Mexico, Albuquerque, NM 87106, USA*
[2]*Department of Materials Science & Engineering, University of Michigan, Ann Arbor, MI 48109, USA*
[3]*Applied Physics Program, University of Michigan, Ann Arbor, MI 48109, USA*
[4]*Department of Electrical and Computer Engineering, The Ohio State University, Columbus, OH 43210, USA*
[5]*Lumileds LLC, San Jose, CA 95131, USA*
[6]*Department of Materials Science and Engineering, The Ohio State University, Columbus, OH 43210, USA*

*Electronic mail:* xuefengli@ucsb.edu

**Scaling Exponent of SRH and Radiative Recombination**

The SRH recombination rate is proportional to the following overlap of the electron and hole wave functions, $R_{SRH} \propto \int dz\, n_T(z) \frac{c_n \psi_e(z) \times c_p \psi_h(z)}{c_n \psi_e(z) + c_p \psi_h(z)}$, where $c_p$ and $c_n$ are the defect carrier capture coefficients [1]. To evaluate the influence of the spatial position of the traps, we set $n_T(z) = \delta(z - Z_D)$. Then,

$$R_{SRH} \propto \frac{c_n \psi_e(Z_D) \times c_p \psi_h(Z_D)}{c_n \psi_e(Z_D) + c_p \psi_h(Z_D)} \quad (1)$$

Likewise, the radiative rate is proportional to the following overlap, $R_r \propto \left| \int dz\, \psi_e^*(z) \psi_h(z) \right|^2$. In a polar quantum well, we can approximate the wave function as $\psi_e(z) \sim \exp\left(\frac{z-L}{\xi_e}\right)$ and $\psi_e(z) \sim \exp\left(-\frac{L}{\xi_h}\right)$, where $\xi$ is the spatial extent of the wave function and $L$ is the quantum well width. Since $\xi_h \ll \xi_e$ in III-nitride quantum wells [2], we can simplify the radiative rate as follows,

$$R_r \propto |\psi_e(0)|^2 \sim \exp\left(-\frac{2L}{\xi_e}\right) \quad (2)$$

Similarly, we simplify the SRH rate as follows. Since we would like show that defects localized near $z = L$ can give rise to a super-linear relation between $R_{SRH}$ and $R_r$, we set $Z_D \approx L$ and substitute the exponential approximation for the wave functions, which yields,

$$R_{SRH} \propto \frac{\exp\left(-\frac{2L}{\xi_h}\right)}{1 + \frac{c_p}{c_n} \exp\left(-\frac{2L}{\xi_h}\right)} \quad (3)$$

Recognizing that the localization length is inversely proportional to the effective mass according to the Agmon relation [2], we write the following relation, $\xi_h \approx \sqrt{\frac{m_e}{m_h}} \xi_e$. Combining this result with equation (2) and substituting into equation (3), we find,

$$R_{SRH} \propto \frac{\exp\left(-\sqrt{\frac{m_h}{m_e}}\frac{2L}{\xi_e}\right)}{1+\frac{c_p}{c_n}\exp\left(-\sqrt{\frac{m_h}{m_e}}\frac{2L}{\xi_e}\right)} \sim \frac{R_r^{\sqrt{\frac{m_h}{m_e}}}}{1+\frac{c_p}{c_n}R_r^{\sqrt{\frac{m_h}{m_e}}}} \quad (4)$$

For illustrative purposes, we set $c_p \ll c_n$. In this limit, the expression can be further simplified as,

$$R_{SRH} \propto R_r^{\sqrt{\frac{m_h}{m_e}}} \quad (5)$$

Since $m_h \approx 10\, m_e$ in InGaN, this shows that the SRH rate can scale superlinearly with the radiative recombination rate. Hence, in principle, the SRH rate can increase more rapidly than the radiative recombination rate as the quantum well thickness decreases. While beyond the scope of the present work, this effect should be further investigated in future works.

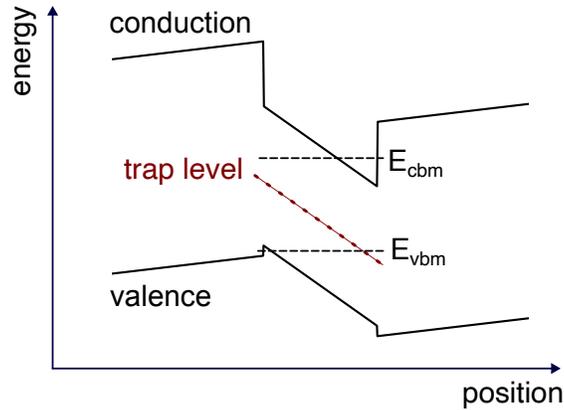

FIG. S1. Influence of polarization fields on trap levels. LEDs comprise polar quantum wells that exhibit strong polarization fields, which can shift the energy of defect levels, relative to the VBM and CBM.